\documentclass[%
 aip,
 amsmath,amssymb,
 reprint,%
]{revtex4-1}

\usepackage{graphicx}% Include figure files
\usepackage{dcolumn}% Align table columns on decimal point
\usepackage{bm}% bold math
\usepackage[utf8]{inputenc}
\usepackage[T1]{fontenc}
\usepackage{mathptmx}

\usepackage{amsmath}  % needed for \tfrac, \bmatrix, etc.
\usepackage{amsfonts} % needed for bold Greek, Fraktur, and blackboard bold
\usepackage{graphicx} % needed for figures

\begin{document}

% Be sure to use the \title, \author, \affiliation, and \abstract macros
% to format your title page.  Don't use lower-level macros to  manually
% adjust the fonts and centering.

\author{Gerard McCaul}
\email{gmccaul@tulane.edu}
\affiliation{Department of Physics, Tulane University, New Orleans, LA 70118, USA}
\author{Andreas Mershin} 
\affiliation{Center for Bits and Atoms, Massachusetts Institute of Technology, Cambridge, MA 02139, USA}
\author{Denys I. Bondar}
\affiliation{Department of Physics, Tulane University, New Orleans, LA 70118, USA}

% \author{anon A}

% \author{anon B} 

% \author{anon C}

\title{Diffusion Fails to Make a Stink}
% In a long title you can use \\ to force a line break at a certain location.

%\author{Daniel V. Schroeder}
%\email{dschroeder@weber.edu} % optional
%\altaffiliation[permanent address: ]{101 Main Street, 
%  Anytown, USA} % optional second address
%% If there were a second author at the same address, we would put another 
%% \author{} statement here.  Don't combine multiple authors in a single
%% \author statement.
%\affiliation{Department of Physics, Weber State University, Ogden, UT 84408-2508}
%% Please provide a full mailing address here.
%
%\author{David P. Jackson}
%\email{ajp@dickinson.edu}
%\affiliation{Department of Physics, Dickinson College, Carlisle, PA 17013}

% See the REVTeX documentation for more examples of author and affiliation lists.

\date{\today}

\begin{abstract}
In this work we consider the question of whether a simple diffusive model can explain the scent tracking behaviors found in nature. For such tracking to occur, both the concentration of a scent and its gradient must be above some threshold. Applying these conditions to the solutions of various diffusion equations, we find that {the steady state of a} purely diffusive model cannot simultaneously satisfy the tracking conditions when parameters are in the experimentally observed range. This demonstrates the necessity of modelling odor dispersal with full fluid dynamics, where non-linear phenomena such as turbulence play a critical role.

\end{abstract}
% AJP requires an abstract for all regular article submissions.
% Abstracts are optional for submissions to the "Notes and Discussions" section.

\maketitle
\section{Introduction}

We live in a universe that not only obeys mathematical laws, but on a fundamental level appears determined to keep those laws comprehensible \cite{Wigner}. The achievements of physics in the three centuries since the publication of Newton's \emph{Principia Mathematica} \cite{newton1999the} are largely due to this inexplicable contingency. The predictive power of mathematical methods has spurred its adoption in fields as diverse as social science \cite{weidlich2006sociodynamics} and history \cite{Vugt_2017}. A particular beneficiary in the spread of mathematical modelling has been biology \cite{Cohen2004}, which has its origins in Schr{\"o}dinger's analysis of living beings as reverse entropy machines \cite{Schrodinger}. Today, mathematical treatments of biological processes abound, modelling everything from epidemic networks \cite{Hethcote2000, Ulrich2006} to biochemical switches \cite{Hernansaiz-Ballesteros2018}, as well as illuminating deep parallels between the processes driving both molecular biology and silicon computing \cite{Dalchau2018}.  

One of the most natural applications of mathematical modelling is to understand the sensory faculties through which we experience the world. Newton's use of a bodkin to deform the back of his eyeball \cite{darrigol2012a, bryson2003a} was one of many experiments performed to confirm his theory of optics \cite{newton2012opticks, NewtonOptics1,NewtonOptics2}. Indeed, the experience of both sight and sound have been extensively contextualised by the  mathematics of optics \cite{optics1,optics2, Luneburg1947, optics3} and acoustics  \cite{Rutherford1886, hearingbar, Gabor1947, hearingbeats}. In contrast to this, simple models which adequately describe the phenomenological experience of smell are strangely lacking, belying the important role olfaction plays in our perception of the world \cite{sell2019fundamentals}. A robust model describing scent dispersal is of some importance, as olfaction has the potential to be used in the early diagnosis  \cite{Bijland2013} of infections \cite{Bomers2012} and cancers \cite{Buszewski2012, Willis712,Else2020}. In fact, recent work using canine olfaction to train neural networks in the early detection of prostate cancers \cite{Mershin} suggests that future technologies will rely on a better understanding of our sense of smell.  

In the face of these developments, it seems timely to revisit the mechanism of odorant dispersal, and examine the consequences of modeling it via diffusive processes. Here we explore the consequences of using the mathematics of diffusion to describe the dynamics of odorants. In particular, we wish to understand whether such simple models can account for the capacity of organisms to not only detect odors, but to track them to their origin. In previous work, the process of olfaction \emph{inside} the nasal cavity has been modeled with diffusion \cite{Hahn1994}, but the question of whether purely diffusive processes can lead to spatial distributions of scent concentration that enable odor tracking has not been considered.    

The phenomenon of diffusion has been known and described for millennia, an early example being Pliny the Elder's observation that it was the process of diffusion that gave roman cement its strength \cite{pliny1991natural, cement}.  Diffusion equations have been applied to scenarios as diverse as predicting a gambler's casino winnings \cite{gambler} to baking a cake \cite{diffusioncake}. The behavior described by the diffusion equation is the \emph{random spread} of substances \cite{diffusion1,diffusion2}, with its principal virtue being that it is described by well-understood partial differential equations whose solutions can often be obtained analytically. It is therefore a natural candidate for modelling random-motion transport such as (appropriately in 2020) the spread of viral infections \cite{covid} or the dispersal of a gaseous substance  such as an odorant. 

The rest of this paper is organized as follows - in Sec.\ref{sec:Modelling}, we introduce the diffusion equation, and the conditions required of its solution to both detect and track an odor. Sec.\ref{sec:simple} solves the simplest case of diffusion, which applies in scenarios such as a drop of blood diffusing in water. This model is extended in Sec.\ref{sec:source} to include both source and decay terms, which describes e.g. a pollinating flower. Finally Sec.\ref{sec:conclusions} discusses the results presented in previous sections, which find that the distributions which solve the diffusion equation cannot be reconciled with experiential and empirical realities. Ultimately the processes that enable our sense of smell cannot be captured by a simple phenomenological description {of time-independent spatial distributions}, and models for the olfactory sense must account for the non-linear \cite{nonlinearode} dispersal of odor caused by secondary phenomena such as turbulence.

\section{Modelling Odor Tracking  with Diffusion \label{sec:Modelling}}
We wish to answer the question of whether a simple mathematical model can capture the phenomenon of tracking a scent. We know from experience that it is possible to trace the source of an odorant, so any physical model of the dispersal of odors must capture this fact. The natural candidate model for this is the \emph{diffusion equation}, which in its most basic  (one-dimensional) form is given by \cite{riley2006mathematical} 
\begin{equation}
\frac{\partial C\left(x,t\right)}{\partial t}-D\frac{\partial^{2}C\left(x,t\right)}{\partial x^{2}}=0
\label{eq:Diffusion}
\end{equation}
where $C(x,t)$ is the concentration of the diffusing substance, and $D$ is the diffusion constant determined by the microscopic dynamics of the system.  For the sake of notational simplicity, all diffusion equations presented in this manuscript will be 1D. An extension to 3D will not change any conclusions that can be drawn from the 1D case, as typically the spatial variables in a diffusion equation are separable, so that a full 3D solution to the equation will simply be the product of the 1D equations (provided the 3D initial condition is the product
of 1D conditions). 

\begin{figure}[!h]
\begin{center}
\includegraphics[width=1\linewidth]{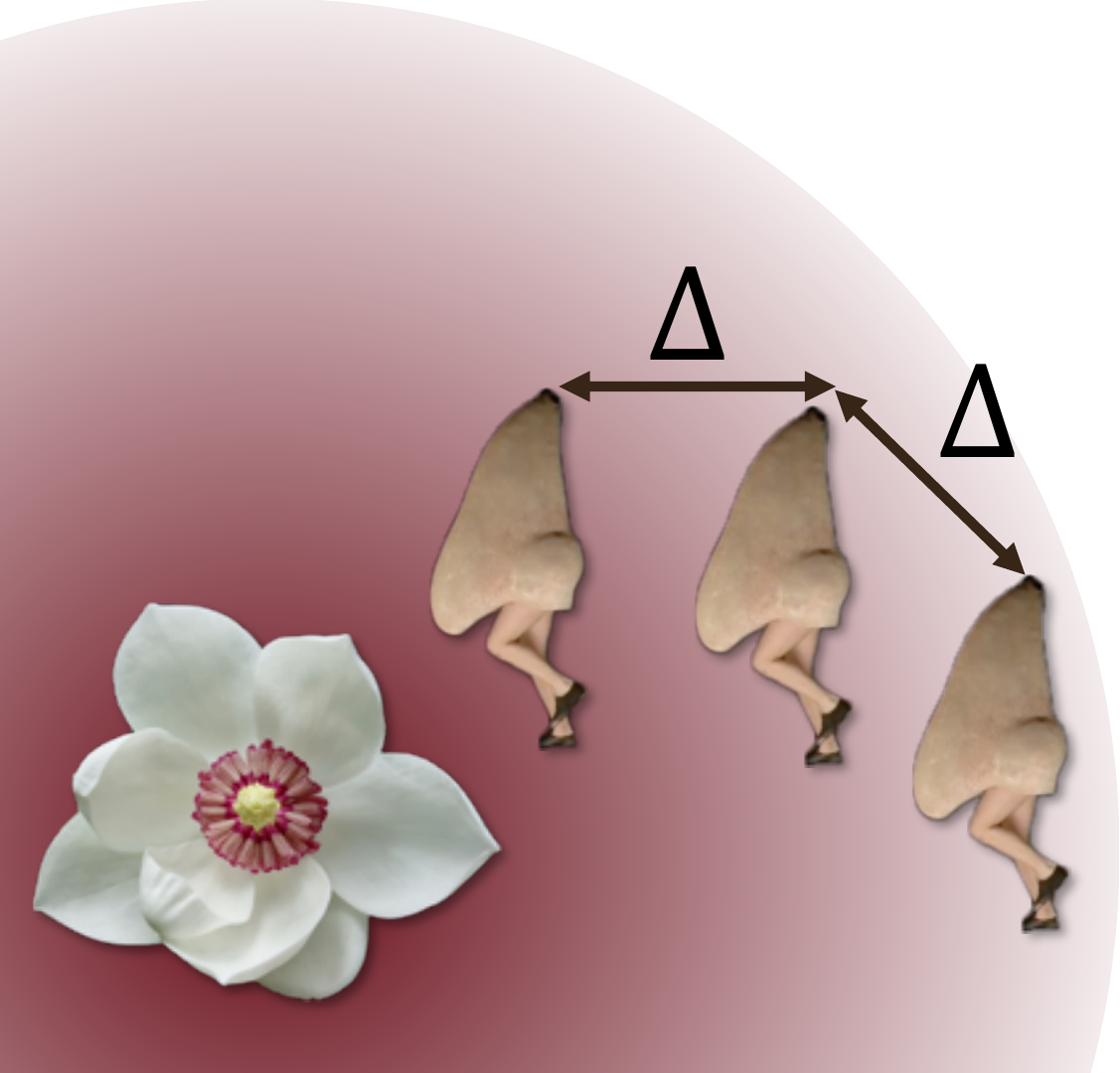}
\end{center}
\caption{{\bf Tracking Odors} In order to trace an odor to its source, one sniffs at different locations separated by $\Delta$. If the concentration gradient is sufficiently large, it is possible to determine if one is closer or further from the origin of the smell. Image of the walking nose comes from the Royal Opera House's production of Shostakovich's `The Nose'.}
\label{fig:odortrack}
\end{figure}

If an odorant is diffusing according to Eq.\eqref{eq:Diffusion} or its generalizations, there are two prerequisites for an organism to track the odor to its source. First, the odorant must be detectable, and therefore its concentration at the position of the tracker should exceed a given threshold or \emph{Limit Of Detection} (LOD) \cite{Sell2006, NICOLAS2004384}. Additionally,  one must be able to distinguish relative concentrations of the odorant at different positions in order to be able to follow the concentration gradient to its source. Fig. \ref{fig:odortrack} sketches the method by which odors are tracked, with the organism sniffing at different locations (separated by a length $\Delta$) in order to find the concentration gradient that determines which direction to travel in. 

We can express these conditions for tracking an odorant with two equations
\begin{align}
    C(x)&>C_T,
    \label{eq:threshold} \\
\frac{C(x)}{C(x+\Delta)}&>R
\label{eq:sensitivity}
\end{align}
where $C(x)$ is the {spatial} distribution of odor concentration at some time, $C_T$ is the LOD concentration, and $R$ characterises the sensitivity to the concentration gradient when smelling at positions $x$ and $x+\Delta$ {(where $x+\Delta$ is further from the scent origin)}.   

The biological mechanisms of olfaction determine both $C_T$ and $R$, and can be estimated from empirical results. While the LOD varies greatly across the range of odorants and olfactory receptors, the lowest observed thresholds are on the order of 1 part per billion (ppb) \cite{Ding2014}. Estimating $R$ is more difficult, but a recent study in mice demonstrated that a 2-fold increase in concentration between inhalations was sufficient to trigger a cellular response in the olfactory bulb \cite{Parabucki2019}. { Furthermore, comparative studies have demonstrated similar perceptual capabilities between humans and rodents \cite{10.1093/chemse/bjt057}}.  We therefore assume that in order to track an odor, $R\approx 2$. Values of $\Delta$ will naturally depend on the size of the organism and its frequency of inhalation, but unless otherwise stated we will assume $\Delta =1{\rm m}$.  

Having established the basic diffusion model and the criteria necessary for it to reflect reality, we now examine under what conditions the solutions to diffusion equations are able to satisfy Eqs.(\ref{eq:threshold},\ref{eq:sensitivity}).

\section{The Homeopathic Shark \label{sec:simple}}

 Popular myth insists that the predatory senses of sharks allow them to detect a drop of its victim's blood from a mile away, although in reality the volumetric limit of sharks' olfactory detection is about that of a small swimming pool \cite{Meredith2010}. {While in general phenomena such as Rayleigh-Taylor instabilities \cite{Taylor1,Taylor2,Taylor3,Taylor4} can lead to mixing at the fluid interface, in the current case the similar density of blood and water permits such effects to be neglected}. The diffusion of a drop of blood in water is therefore precisely the type of scenario in which Eq.\eqref{eq:Diffusion} can be expected to apply. To test whether this model can be reconciled to reality, we first calculate the predicted maximum distance $x_{\rm max}$ from which the blood can be detected. 

In order to find $C(x,t)$, we stipulate that the mass $M$ of blood is initially described by $C(x,0)=M\delta(x)$. While many methods exist to solve Eq.\eqref{eq:Diffusion}, the most direct is to consider the \emph{Fourier transform} of the concentration  \cite{evans2010partial}:
\begin{equation}
\tilde{C}(k,t)=\mathcal{F}[C(x,t)]=\int^\infty_{-\infty} {\rm d}x\  {\rm e}^{-i k x}C(x,t). 
\end{equation}
Taking the time derivative and substituting in the diffusion equation we find 
\begin{equation}
    \frac{\partial \tilde{C}(k,t)}{\partial t}=D\int^\infty_{-\infty} {\rm d}x\  {\rm e}^{-i k x}\frac{\partial^2 C(x,t)}{\partial x^2}.
\end{equation}
The key to solving this equation is to integrate the right hand side by parts twice. If the boundary conditions are such that both the concentration and its gradient vanish at infinity, then the integration by parts results in  
\begin{equation}
    \frac{\partial \tilde{C}(k,t)}{\partial t}=-D k^2 \tilde{C}(k,t).
\end{equation}
This equation has the solution 
\begin{equation}
\tilde{C}(k,t)= \tilde{f}(k){\rm e}^{-D k^2 t} 
\end{equation}
where the function $\tilde{f}(k)$ corresponds to the Fourier transform of the initial condition. In this case (where $C(x,0)=M\delta(x)$),  $\tilde{f}(k)=M$. 
The last step is to perform the inverse Fourier transform to recover the solution
\begin{equation}
    C(x,t)=\mathcal{F}^{-1}[\tilde{C}(k,t)]=\frac{M}{2\pi}\int^\infty_{-\infty} {\rm d}k \ {\rm e}^{-Dk^2t+ikx}.
    \label{eq:DiffusionInverseFourier}
\end{equation}
The integral on the right hand side is a Gaussian integral, {and can be solved using the standard procedure of completing the square in the integrand exponent \cite{riley2006mathematical,kantorovich2016mathematics}. The final solution to Eq.(\ref{eq:Diffusion}) is then}
\begin{equation}
C\left(x,t\right)=\frac{M}{2\pi}{\rm e}^{-\frac{x^2}{4Dt}}\int^\infty_{-\infty} {\rm d}k \ {\rm e}^{-Dk^2t} =\frac{M}{\sqrt{4\pi Dt}}{\rm e}^{-\frac{x^{2}}{4Dt}}\label{eq:blood-distribution}.
\end{equation}

This expression for the concentration is dependent on both time and
space, however for our purposes we wish to understand the threshold
sensitivity with respect to distance. To that end, we consider the
concentration $C^{*}\left(x\right)$, \emph{which describes the highest
concentration at each point in space across all of time.} This is derived
by calculating the time which maximises $C(x,t)$ at each
point in $x$:
\begin{align}
\frac{\partial C\left(x,t\right)}{\partial t}&=\frac{M}{\sqrt{4\pi Dt}}{\rm e}^{-\frac{x^{2}}{4Dt}}\left(\frac{x^{2}}{4Dt^{2}}-\frac{1}{2t}\right),
\label{eq:simplesolution} \\
\frac{\partial C\left(x,t^*\right)}{\partial t}&=0\implies t^{*}=\frac{x^{2}}{2D}.
\end{align}
 Using this, we have
\begin{equation}
C^{*}\left(x\right)=C\left(x,t^{*}\right)=\frac{M}{\sqrt{2\pi e}x}, \label{eq:bloodsteady}
\end{equation}
 where $e$ is Euler's number. This distribution represents a ``best-case'' scenario, where one happens to be in place at the right time for the concentration to be at its maximum.
Interestingly, {while the time of maximum concentration depends on $D$, the concentration itself} is entirely insensitive to the microscopic dynamics governing $D$ - the maximum distance a transient scent can be detected is the same whether the shark is swimming through water or treacle! 

\begin{figure}[!h]
\begin{center}
  \includegraphics[width=\linewidth]{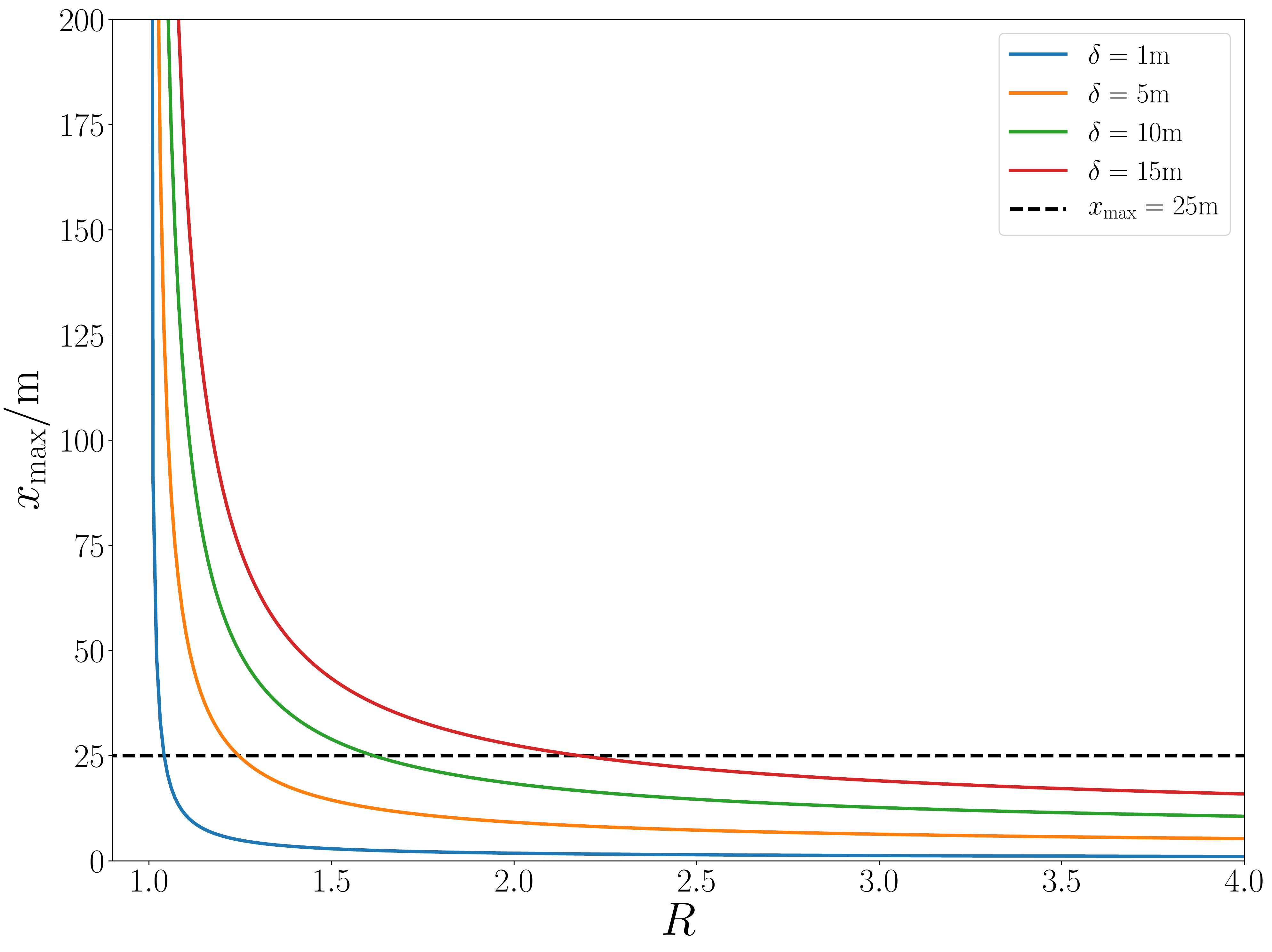}
\end{center}
\caption{{\bf Gradient Sensitivity:} The maximum trackable distance depends strongly on both the minimum gradient sensitivity $R$ and the spacing between inhalation $\Delta$. In order to obtain an $x_{\rm max}$ comparable with that associated with the LOD using $R=2$, $\Delta$ must be on the order of $x_{\rm max}$.}
\label{fig:Rsense}
\end{figure}

The threshold detection distance $x_{\rm max}$ can be estimated from Eq.\eqref{eq:threshold} using $x_{\rm max}= \frac{M}{\sqrt{2\pi e}C_T}$. For a mass of blood $M=1\rm{g}$ and an estimated LOD of $C_T = 1 \rm{ppb} \sim 1 \mu\rm{gm^{-3}}$. As we are working in one dimension we take the cubic root of this threshold to obtain $x_{\rm{max}}\approx 25\rm{m}$.
While this seems a believable threshold for detection distances, is it possible to \emph{track} the source of the odor from this distance? Returning to Eq.\eqref{eq:simplesolution}, the ratio when the concentration is maximal at $x$ is
\begin{equation}
\frac{C(x,t^{*})}{C(x+\Delta,t^{*})}=\exp\left(\frac{\Delta}{x}+\frac{\Delta^{2}}{2x^{2}}\right).
\end{equation}
Note that this expression assumes that the timescale over which the concentration changes is much slower than the time between inhalations, hence we compare the concentrations at $x$ and $x+\Delta$ at the \emph{same} time $t^*$. Setting $\frac{C(x_{\max},t^{*})}{C(x_{\max}+\Delta,t^{*})}=R$, we obtain
\begin{equation}
 x_{\rm max}=\frac{\Delta\left(1+\sqrt{1+2\ln\left(R\right)}\right)}{2\ln\left(R\right)}.
\end{equation}
For the sensitivity $R= 2$, $x_{\rm max} \approx 1.8\Delta$. {This means that} in order to track the scent, the shark has to start on the order of $\Delta$ away from it.  Fig.\ref{fig:Rsense} shows that to obtain a gradient sensitivity at comparable distances to the LOD distance for $\Delta=1$m  would require $R\approx 1.04$. Even in this idealised scenario, the possibility of the shark being able to distinguish and act on a 4\% increase in concentration is remote. This suggests that the diffusion model is doing a poor job capturing the real physics of the blood dispersion, and/or the shark's ability to sense a gradient is somehow improved when odorants are at homeopathically low concentrations. {Here we see the first example of a theme which will recur in later sections - diffusive processes generate odorant gradients which are too shallow to follow when one is close to $x_{\rm max}$.

An important caveat should be made to this and later results, namely that the odor tracking strategy we have considered depends purely on the \emph{spatial} concentration distribution at a particular moment in time. In reality, sharks are just one of a variety of species which rely on scent arrival time to process and perceive odors \cite{GARDINER20101187,Dalal2020}. One might reasonably ask whether this additional capacity could assist in the detection of purely diffusing odors, using a tracking strategy that incorporates memory effects. For the moment, it suffices to note that the timescales in which diffusion operates will be far slower than any time-dependent tracking mechanism. We shall find in Sec. \ref{sec:source} however that the addition of advective processes to diffusion will force us to revisit this assumption.} 

\section{Adding a source \label{sec:source}}
The simple diffusion model in the previous section predicted that at any scent found at the limit of detection would have a concentration gradient too small to realistically track. This is clearly at odds with lived experience, so we now consider a more realistic system, where there is a continuous source of odorant molecules (e.g. a pollinating flower). In this case our diffusion equation is
\begin{equation}
\frac{\partial C\left(x,t\right)}{\partial t}-D\frac{\partial^{2}C\left(x,t\right)}{\partial x^{2}} +KC(x,t)=f\left(x,t\right)
\label{eq:Diffusiondriftdecay}
\end{equation}
where $f\left(x,t\right)$ is a source term describing the product of odorants, and $K$ is a decay constant modelling the finite lifetime of odorant molecules. 

Finding a solution to this equation is more nuanced than the previous example, due to the inhomogeneous term $f(x,t)$. For now, let us ignore this term, and consider only the effect of the $KC(x,t)$ decay term. In this case, the same Fourier transform technique can be repeated (using the initial condition $C(x,0)=C_0\delta(x)$), leading to the solution $C_K(x,t)$:
\begin{equation}
C_K(x,t)=\frac{C_0}{\sqrt{4\pi Dt}}{\rm e}^{-\frac{x^{2}}{4Dt}-Kt}.    
\label{eq:HomoSolution}
\end{equation}
This is almost identical to our previous solution, differing only in the addition of a decay term $Kt$ to the exponent. 

Incorporating the source term $f(x,t)$ presents more of a challenge, but it can be overcome with the use of a \emph{Green's function} \cite{rother2017green's}. First we postulate that the solution to the diffusion equation can be expressed as
\begin{equation}
C\left(x,t\right)=\int_{0}^{\infty}{\rm d}\tau\int_{-\infty}^{\infty}{\rm d}\xi\ G\left(x,\xi,t,\tau\right)f\left(\xi,\tau\right),
\label{eq:Greenintegral}
\end{equation}
 where $G$ is known as the Green's function. {Note that for a solution of this form to exist, the right hand side must satisfy the same properties as $C$, namely that the integral of $G\left(x,\xi,t,\tau\right)f\left(\xi,\tau\right)$  under $\tau$ and $\xi$ is an integrable, normalisable function.}  In order for Eq.\eqref{eq:Greenintegral} to satisfy Eq.\eqref{eq:Diffusiondriftdecay}, $G$ must itself satisfy:
 \begin{align}
\frac{\partial G\left(x,\xi,t,\tau\right)}{\partial t}-&D\frac{\partial^{2}G\left(x,\xi,t,\tau\right)}{\partial x^{2}} \notag \\ +&KG\left(x,\xi,t,\tau\right)= \delta(t-\tau) \delta(x-\xi)
\label{eq:DiffusionGreenFunction}
\end{align}
Note that the consistency of Eq.\eqref{eq:Greenintegral}  with Eq.\eqref{eq:Diffusiondriftdecay} can be easily verified by substituting  Eq.\eqref{eq:DiffusionGreenFunction} into it.

At first blush, this Green's function equation looks no easier to solve than the original diffusion equation for $C(x,t)$. Crucially however, the inhomogeneous forcing term $f(x,t)$ has been replaced by a product of delta functions which may be analytically Fourier transformed. Performing this transformation on $x$, we find
\begin{equation}
\frac{\partial \tilde{G}\left(k,\xi,t,\tau\right)}{\partial t}-(Dk^2-K)\tilde{G}\left(k,\xi,t,\tau\right)={\rm e}^{-ik\xi}\delta(t-\tau).
\label{eq:GreenFouriereq}
\end{equation}
We can bring the entirety of the left hand side of this expression under the derivative with the use of an \emph{integrating factor} \cite{kantorovich2016mathematics1}. In this case, we observe that
% \begin{align}
% \frac{\partial {\rm e}^{-(Dk^2-K)t}\tilde{G}\left(k,\xi,t,\tau\right)}{\partial t}&={\rm e}^{-(Dk^2-K)t}\bigg[\frac{\partial \tilde{G}\left(k,\xi,t,\tau\right)}{\partial t}
% \notag \\  &-(Dk^2-K)\tilde{G}\left(k,\xi,t,\tau\right)\bigg],   
% \end{align}
\begin{align}
\frac{\partial}{\partial t}\left({\rm e}^{-(Dk^2-K)t}\tilde{G}\left(k,\xi,t,\tau\right)\right)&={\rm e}^{-(Dk^2-K)t}\bigg[\frac{\partial \tilde{G}\left(k,\xi,t,\tau\right)}{\partial t}
\notag \\  &-(Dk^2-K)\tilde{G}\left(k,\xi,t,\tau\right)\bigg],   
\end{align}
% \begin{align}
% \frac{\partial {\rm e}^{-(Dk^2-K)t}\tilde{G}}{\partial t}&={\rm e}^{-(Dk^2-K)t}\bigg[\frac{\partial \tilde{G}}{\partial t}
% \notag \\  &-(Dk^2-K)\tilde{G}\bigg],   
% \end{align}
which can be substituted into Eq.\eqref{eq:GreenFouriereq} to obtain
\begin{equation}
\frac{\partial}{\partial t}\left({\rm e}^{-(Dk^2-K)t}\tilde{G}\left(k,\xi,t,\tau\right)\right)={\rm e}^{(Dk^2-K)t}{\rm e}^{-ik\xi}\delta(t-\tau).
\end{equation}
Integrating both sides (together with the initial condition $C(x,0)=G(x,\xi,0,\tau)=0$) yields the Green's function in $k$ space:
\begin{equation}
\tilde{G}\left(k,\xi,t,\tau\right)= \theta_H(t-\tau){\rm e}^{-(Dk^2-K)(t-\tau)}{\rm e}^{-ik\xi}
\end{equation}
where $\theta_H(t-\tau)$ is the Heaviside step function. The inverse Fourier transform of this function is once again a Gaussian integral, and can be solved for in an identical manner to Eq.\eqref{eq:DiffusionInverseFourier}. Performing this integral, we find 
\begin{equation}
    G(x,\xi,t,\tau)=\theta_H(t-\tau) \frac{1}{\sqrt{4\pi D(t-\tau)}}{\rm e}^{-\frac{(x-\xi)^{2}}{4D(t-\tau)}-K(t-\tau)}. 
\end{equation}
Note that this Green's function for an inhomogeneous diffusion equation with homogeneous initial conditions is essentially the solution $C_K$ given in Eq.\eqref{eq:HomoSolution} to the homogeneous equation with an inhomogeneous initial condition! This surprising result is an example of \emph{Duhamel's principle} \cite{tikhonov1990equations}, which states that the source term can be viewed as the \emph{initial condition} for a new homogeneous equation starting at each point in time and space. The full solution will then be the integration of each of these homogeneous equations over space and time, exactly as suggested by Eq.\eqref{eq:Greenintegral}. From this perspective, it is no surprise that the Green's function is so intimately connected to the unforced solution. 

Equipped with the Green's function, we are finally ready to tackle Eq.\eqref{eq:Greenintegral}. Naturally, this equation is only analytically solvable when $f\left(x,t\right)$
is of a specific form. We shall therefore assume flower's pollen production is time independent and model it as a point source
$f\left(x,t\right)$=$J\delta\left(x\right)$. In this case the concentration
is given by
\begin{equation}
C\left(x,t\right)=  \frac{J}{\sqrt{4\pi D}}\int_{0}^{t}{\rm d}\tau\ \frac{1}{\sqrt{\tau}}{\rm e}^{-\frac{x^{2}}{4D\tau}-K\tau}.
\end{equation}
Now while it is possible to directly integrate this expression, the result is a collection of error functions \cite{riley_hobson_bence_2002}. For both practical and aesthetic reasons, we therefore consider the \emph{steady state} of this distribution $C_s(x)$:
\begin{equation}
    \lim_{t\to\infty} C\left(x,t\right)= C_s(x)= \frac{J}{\sqrt{4\pi D}}\int_{0}^{\infty}{\rm d}\tau\ \frac{1}{\sqrt{\tau}}{\rm e}^{-\frac{x^{2}}{4D\tau}-K\tau}.
    \label{eq:steadystateinit}
\end{equation}
This integral initially appears unlike those we have previously encountered, but ultimately we will find that this is yet another Gaussian integral in deep cover. To begin this process, we make the substitution $t=\sqrt{\tau}$:
\begin{align}
\int_{0}^{\infty}{\rm d}\tau\ \frac{1}{\sqrt{\tau}}{\rm e}^{-\frac{x^{2}}{4D\tau}-K\tau} &=2\int_{0}^{\infty}{\rm d}t\ {\rm e}^{-\frac{x^{2}}{4Dt^2}-Kt^2} \notag \\ &=\int_{-\infty}^{\infty}{\rm d}t\ {\rm e}^{-\frac{x^{2}}{4Dt^2}-Kt^2}
\end{align}
where the last equality exploits the even nature of the integrand. At this point we perform another completion of the square, rearranging the exponent to be
\begin{equation}
    -\frac{x^{2}}{4Dt^2}-Kt^2=-\left(\sqrt{K}t-\frac{|x|}{2\sqrt{D}t}\right)^2 -\sqrt{\frac{K}{D}}|x|.
\end{equation}
Combining this with the substitution $t\to \left(\frac{|x|}{2kD}\right)^{\frac{1}{4}}t$, we can express the steady state concentration as:
\begin{equation}
C_s(x)= \frac{J{\rm e}^{-\sqrt{\frac{K}{D}}|x|}}{\sqrt{4\pi D}}\left(\frac{|x|}{2KD}\right)^{\frac{1}{4}}\int_{-\infty}^{\infty}{\rm d}t \  {\rm e}^{-\sqrt{\frac{|x|K}{2D}}\left(t-\frac{1}{t}\right)^2}.
\label{eq:steadystateintermediate}
\end{equation} 
It may appear that the integral in this expression is no closer to being solved than in Eq.\eqref{eq:steadystateinit}, but we can exploit a useful property of definite integrals to finish the job.

Consider a general integral of the form $\int_{-\infty}^{\infty}{\rm d}x \  f(y)$, where $y=x-\frac{1}{x}$. Solving the latter expression, we see that $x$ has two possible branches,
\begin{equation}
    x_\pm = \frac{1}{2}\left(y \pm \sqrt{y^2+4}\right).
\end{equation}
Using this, we can split the integral into a term integrating along each branch of $x$:
\begin{align}
    \int_{-\infty}^{\infty}{\rm d}x \  f(y) &=\int_{-\infty}^{0_-}{\rm d}x_-\  f(y) +\int_{0_+}^{\infty}{\rm d}x_+ \ f(y) \notag \\ &= \int_{-\infty}^{\infty}{\rm d}y  \ \left(\frac{{\rm d}x_-}{{\rm d}y}+\frac{{\rm d}x_+}{{\rm d}y}\right) f(y).
\end{align}
Evaluating the derivatives, we find $\left(\frac{{\rm d}x_-}{{\rm d}y}+\frac{{\rm d}x_+}{{\rm d}y}\right)=1$, and therefore 
\begin{equation}
    \int_{-\infty}^{\infty}{\rm d}x \  f(y)=\int_{-\infty}^{\infty}{\rm d}y \  f(y).
\label{eq:Glassermaster}
\end{equation}
This remarkable equality is the \emph{Cauchy-Schl{\"o}mlich transformation}\cite{Cauchy, Amdeberhan2019}, and its generalization to both finite integration limits and a large class of substitutions $y(x)$ is known as \emph{Glasser's master theorem}\cite{Glasser}. 

Equipped with Eq.\eqref{eq:Glassermaster}, we can immediately recognise Eq.\eqref{eq:steadystateintermediate} as a Gaussian integral, and evaluate it to obtain our final result
\begin{equation}
C_S(x)=\frac{J{\rm e}^{-\lambda\left|x\right|}}{2\sqrt{DK}},
\label{eq:steadystateconc}
\end{equation}
where $\lambda=\sqrt{K/D}$ is the characteristic length scale of the system. {Physically, this parameter describes the competition between diffusion and decay. As we shall see, for large $\lambda$ decay dominates the dynamics, quickly forcing odorants down to undetectable concentrations. Conversely for small $\lambda$, diffusion is the principal process, spreading the odorant to the extent that the gradient of the steady state is too shallow to track.}   

\begin{figure}[h!]
\begin{center}\includegraphics[width=\linewidth]{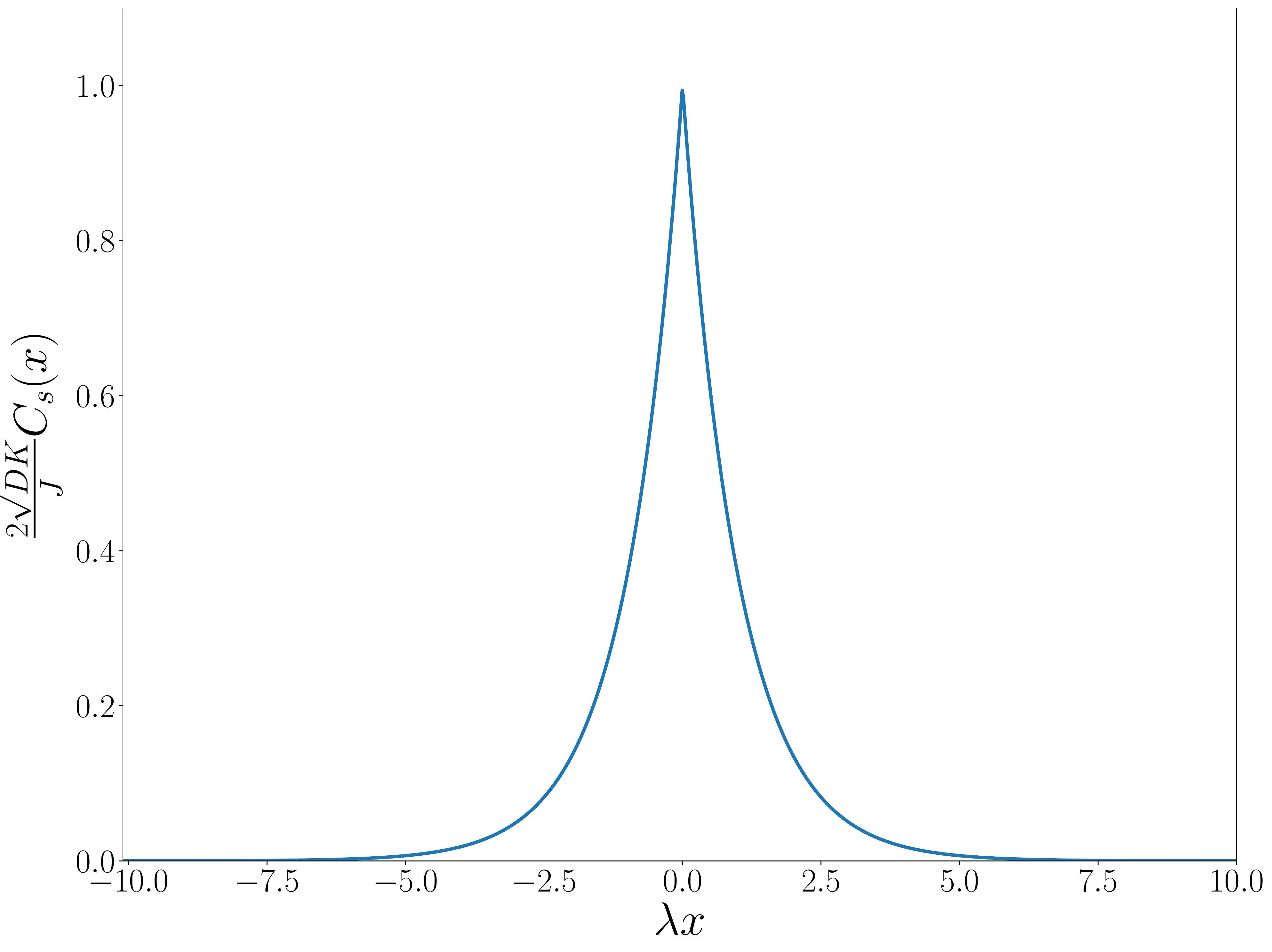}\end{center}\caption{ \textbf{Steady state solution for a diffusing system with both source and decay}: While the source term $J$ only determines the maximum concentration at the origin, the degree of exponential fall-off is strongly dependent on $\lambda=\sqrt{\frac{K}{D}}$.}
\label{fig:steadystate}
\end{figure}

Having finally found our steady state distribution (plotted in Fig.\ref{fig:steadystate}), we can return to the original question of whether this model admits the possibility of odorant tracking. Substituting $C_s(x)$ into Eqs.(\ref{eq:threshold},\ref{eq:sensitivity}), we obtain our maximum distances for surpassing the LOD concentration
\begin{equation}
    x_{\rm max}=\lambda^{-1}\ln{\left(\frac{J}{2C_T\sqrt{DK}}\right)}
    \label{eq:LOD_max}
\end{equation}
and the gradient sensitivity threshold
\begin{equation}
    R_{\rm min}={\rm e}^{\lambda \Delta}.
\end{equation}

Immediately we see that both of these thresholds are most strongly dependent on the characteristic length scale $\lambda$. For the LOD distance, the presence of a logarithm means that even if the LOD were lowered by an order of magnitude, $C_T\to \frac{1}{10}C_T$, the change in $x_{\rm max}$ would be only $\approx\frac{2.3}{\lambda}$. This means that for a large detection distance threshold, a \emph{small} $\lambda$ is imperative.

Conversely, in order for concentration gradients to be detectable, we require $R_{\rm min}\approx2$. This means that $\lambda \Delta \approx 1$, but as we have shown, a reasonable LOD threshold distance needs $\lambda \ll 1$, making a concentration gradient impossible to detect without an enormous $\Delta$. 
It's possible to get a sense of the absurd sensitivities required by this model with the insertion of some specific numbers for a given odorant. Linalool is a potent odorant with an LOD of $C_T=3.2\mu{\rm gm^{-3}}$ in air \cite{Elsharif2015}. Its half life due to oxidation is $t_{\frac{1}{2}} \approx1.8\times10^{7}$s \cite{Skold2004}, from which we obtain $K=\frac{\ln\left(2\right)}{t_{\frac{1}{2}}}\approx3.8\times10^{-8}{\rm s}^{-1}$.
To find the diffusion constant, we use the Stokes-Einstein relation\cite{Einstein1905} {(where $k_B$ is the Boltzmann constant and $\eta$ is the fluid's dynamic viscosity\cite{viscosity})}
\begin{equation}
D=\frac{k_{B}T}{6\pi\eta r},
\end{equation}
taking the temperature as $T=288K$ ($15^{\circ}$C, approximately the average surface temperature of Earth). The molar volume of linalool in 178.9 ml${\rm mol^{-1}}$, and if the molecule is modelled as a sphere of radius $r$, we obtain:
\begin{equation}
r=\left(\frac{3}{4\pi N_{A}}\times178.9\times10^{-6}\right)^{1/3}{\rm m}=4.13\times10^{-10}{\rm m},
\end{equation}
where $N_A\approx 6.02\times 10^{23}$ is Avogadro's number. At 288K, $\eta\approx1.8\times10^{-5}{\rm kgm^{-2}s^{-1}}$ and we obtain $D\approx2.83\times10^{-8}{\rm m^{2}s^{-1}}$, which is close to experimentally observed values \cite{Filho2002}.
Using these figures yields $\lambda=\sqrt{\frac{3.8}{2.8}}{\rm m^{-1}}=1.17{\rm m^{-1}}$, which for $\Delta=1{\rm m}$, gives
\begin{equation}
\frac{C_{s}\left(x\right)}{C_{s}\left(x+\Delta\right)}={\rm e}^{1.17}=3.22
\end{equation}
a figure that suggests an easily detectable concentration gradient. 

\begin{figure}
\begin{center}\includegraphics[width=\linewidth]{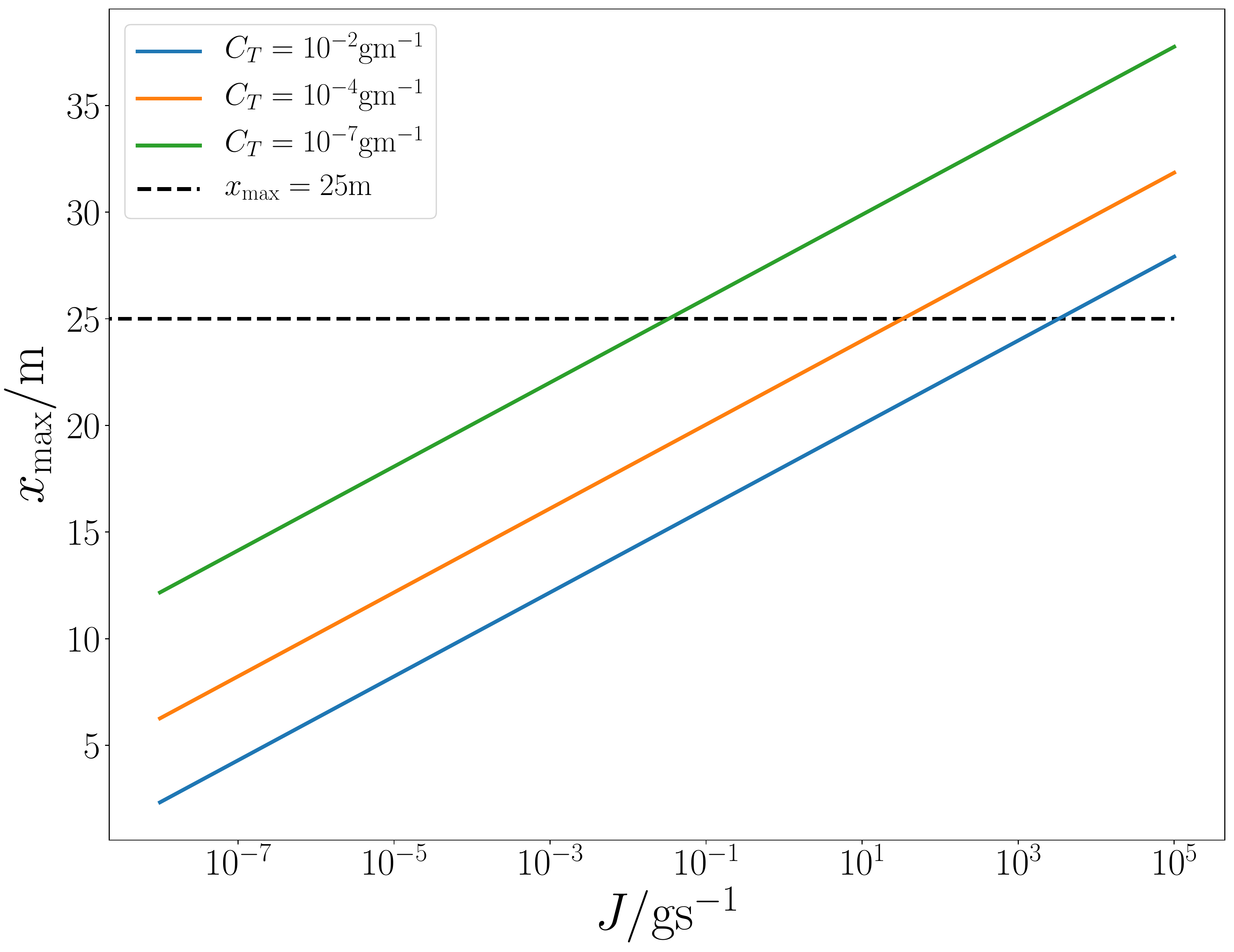}\end{center}\caption{\textbf{Maximum detection distance as a function of source flux}: Using the linalool parameters but varying the LOD threshold, we find that even in the case of $C_T=10^{-7} {\rm gm^{-1}}$ (which corresponds to only $\frac{10^{-21}\times N_A}{154.24} \sim$ 4 molecules per cubic metre), one requires tens of milligrams of odorant being produced each second for detection at $x_{\rm max}=25$ m. At realistic LOD thresholds, the source flux must increase to kilograms per second to reach the same detection distance.}
\label{fig:Jgrowth} 
\end{figure}

As noted before however, a large concentration gradient implies that the LOD distance threshold $x_{\rm max}$ must be very small. Substituting the linalool parameters into Eq.\eqref{eq:steadystateconc} with $x_{\rm max}= 20 {\rm m}$ we find $J= 14 {\rm g s^{-1}}$, i.e. the flower must be producing a mass of odorant on the order of its own weight. If $x_{\rm max}$ is increased to 25m, then the flower must produce \emph{kilograms} of matter every second! Fig.\ref{fig:Jgrowth} shows that even with an artificial lowering of the LOD, unphysically large source fluxes are required.  Once again, the diffusion model is undermined by the brute fact that completely unrealistic numbers are required for odors to be both detectable and trackable.

\subsection*{Adding Drift}
The impossibility of finding physically reasonable parameters which simultaneously satisfy both detection threshold and concentration gradients is due to the \emph{exponential} nature of the concentration distribution, which requires extremely large parameters to ensure that both Eqs.(\ref{eq:threshold},\ref{eq:sensitivity}) hold.  One might question whether the addition of any other dispersal mechanisms can break the steady state's exponential distribution and perhaps save the diffusive model. A natural extension is to add {advection} to the diffusion equation, in order to model the effect of wind currents. The effect of this is to add a term $-v(x,t)\frac{\partial C(x,t)}{\partial x}$ to the right hand side of Eq.\eqref{eq:Diffusiondriftdecay}. For a constant drift $v(x,t)\equiv v$, and $v \gg D,K$ the steady state distribution becomes:
 \begin{equation}
    C_s(x) \approx \left\{
        \begin{array}{ll}
         \frac{J{\rm e}^{-\frac{K}{v}x}}{2v} & x>0, \\ \\
        \frac{J{\rm e}^{\frac{v}{D}x}}{2v}  & x<0.
        \end{array}
    \right.
\end{equation}

Another alternative is to consider a stochastic velocity, with a zero mean $\left<v(t)\right>=0$ and Gaussian auto-correlation $\left<v(t)v(t^\prime)\right>=\sigma \delta(t-t^\prime)$. In this case the average steady state concentration $\left<C_s(x)\right>$ is identical to Eq.\eqref{eq:steadystateconc} with the substitution $D\to D+\sigma$.

In both cases, regardless of whether one adds a constant or stochastic drift the essential problem remains - the steady state distribution remains exponential, and therefore will fail to satisfy one of the two tracking conditions set out in Eqs.(\ref{eq:threshold},\ref{eq:sensitivity}).

{There is however a gap through which these diffusion-advection models might be considered a plausible mechanism for odor tracking. By only considering the steady-state, we leave open the possibility that a time-dependent tracking strategy (as mentioned in Sec. \ref{sec:simple}) may be able to follow the scent to its source during the dynamics' transient period. This will be due precisely to the fact that with the addition of the velocity field $v(x,t)$, the timescale of the odorant dynamics will be greatly reduced. In this case, even if one neglects the heterogeneities that might be induced by a general velocity field, it is possible that the biological processes enabling a time-dependent tracking strategy occupy a timescale compatible with that of the odorant dynamics. In this case, more sophisticated strategies using memory effects could potentially be used to track the diffusion-advection driven odorant distribution.}

\subsection*{Flower Fields}
We have seen that for a single source of scent production, the steady state of the odor distribution does not support tracking, but what about the scenario where a \emph{field} (by which we mean an agricultural plot of land, rather than the algebraic structure often used to represent abstract conditions of space) of flowers is generating odorants? We model this by assuming that a set of $2N+1$ flowers are distributed in the region $x\in [-a,a]$ with a spacing $\Delta_x \ll a$. In this case the distribution will simply be a linear combination of the distributions for individual flowers:
\begin{align}
C_S(x) &=\frac{J}{2\sqrt{DK}}\sum^N_{j=-N} {\rm e}^{-\lambda\left|x-j\Delta_x\right|} \notag \\
&=\frac{J}{2\Delta_x\sqrt{DK}}\sum^N_{j=-N} \Delta_x {\rm e}^{-\lambda\left|x-j\Delta_x\right|},
\end{align}
where in the second equality we have employed a minor algebraic slight of hand so as to approximate the sum as an integral
\begin{equation}
\sum^N_{j=-N} \Delta_x {\rm e}^{-\lambda\left|x-j\Delta_x\right|}\approx \int^a_{-a} {\rm d}y \ {\rm e}^{-\lambda\left|x-y\right|}.
\end{equation}
Note that this is an \emph{approximation} of the sum rather than a limit, so as to obtain a final expression for the distribution while avoiding the issue of taking the limit of $\Delta_x$ outside the sum. With this approximation, the integral can be evaluated analytically (albeit in a piecewise manner),  and the resultant distribution may be seen in Fig.\ref{fig:flowerfield}. 

{Given that for $|x|<a$ one is already within the region of scent production, we will focus our attention on the region $x>a$ (which by symmetry also describes the region $x<-a$). In this case, we have
\begin{equation}
    C_S(x>a) \approx\frac{J{\rm e}^{-\lambda x}}{2\Delta_x\sqrt{DK}}\int^a_{-a} {\rm d}y \ {\rm e}^{\lambda y}=\frac{J\sinh(\lambda a){\rm e}^{-\lambda x}}{\lambda\Delta_x\sqrt{DK}}.
\end{equation}

\begin{figure}
\begin{center}\includegraphics[width=\linewidth]{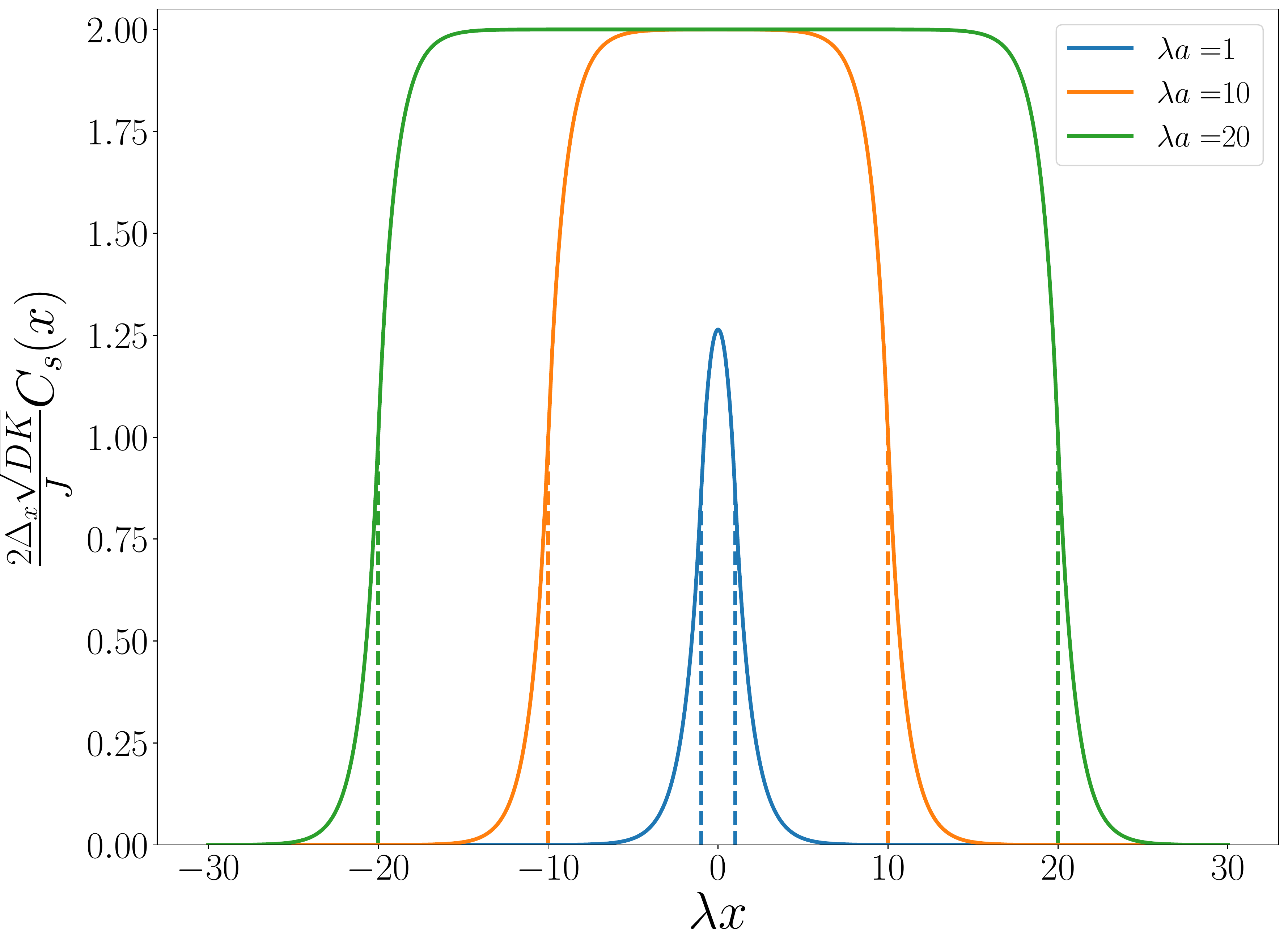}\end{center}\caption{\textbf{Concentrations for a field of of flowers}: The distribution for a field of flowers will rapidly saturate inside the source region $x\in[-a,a] $ (indicated by dashed lines), but outside this region the concentration distribution remains exponential. }
\label{fig:flowerfield} 
\end{figure}

This distribution is identical to Eq.\eqref{eq:steadystateconc} with the substitution $J\to J\frac{2\sinh(\lambda a)}{\lambda\Delta_x}$. One's initial impression might be that this would reduce the necessary value of $J$ for a given LOD threshold by many orders of magnitude, but we must also account for the shift in the scent origin away from $x=0$ to $x=a$. This means that the proper comparison to (for example)  $x_{\rm max}=20$m in the single flower case would be to take $x_{\rm max}=(20 + a)$m here. This extra factor of $a$ will approximately cancel the scaling of $J$ by $2\sinh(\lambda a)$ (for $a>1$). It therefore follows that the effective scaling of $J$ in this region compared to the single flower is only $J \to \frac{J}{\lambda\Delta_x}$. Inserting this into Eq.\eqref{eq:LOD_max}, one sees that for a given $J$, the LOD distance is improved only logarithmically by an additional $\lambda^{-1} \ln(\frac{1}{\lambda \Delta_x})$. Depending on $\Delta_x$, this may improve $x_{\rm max}$ somewhat, but would require extraordinarily dense flower fields to be consistent with the detection distances found in nature. We again stress that these results consider only the steady state distribution, and are therefore subject to the same caveats discussed previously.} 

\section{Discussion \label{sec:conclusions}}
\emph{My Dog has no nose. How does he smell? Terrible.}

In this paper we have considered the implications for olfactory tracking when odorant dispersal is modelled as a purely diffusive process. We find that even under quite general conditions, the steady state distribution of odorants is exponential in its nature. This exponent is characterized by a length scale $\lambda$ whose functional form depends on whether the mechanisms of drift and decay are present. 
The principal result presented here is that in order to track an odor, it is necessary for odor concentrations both to exceed the LOD threshold, and have a sufficiently large gradient to allow the odor to be tracked to its origin. Analysis showed that in exponential models these two requirements are fundamentally incompatible, as large threshold detection distances require small $\lambda$, while detectable concentration gradients need large $\lambda$. Estimates of the size of other parameters necessary to compensate for having an unsuitable $\lambda$ in one of the tracking conditions lead to entirely unphysical figures either in concentration thresholds or source fluxes of odorant molecules. {We emphasise however that these conclusions are drawn on the basis of an odor tracking strategy that incorporates only the spatial information of the odorant distribution, an assumption that holds only when the timescales of the odorant dynamics and scent perception are sufficiently separated.}  

In reality, it is well known that odorants disperse in long, turbulent plumes \cite{Murlis2000,Moore2004} which exhibit extreme fluctuations in concentration on short length scales \cite{plumeexperiment}. It is these spatio-temporal patterns that provide sufficient stimulation to the olfactory senses \cite{Vickers2001}. The underlying dynamics that generate these plumes are a combination of the microscopic diffusive dynamics discussed here, and the turbulent fluid dynamics of the atmosphere, which depend on both the scale and dimensionality of the modeled system \cite{Weissburg2002}. This gives rise to a velocity field $v(x,t)$ that has a highly non-linear spatiotemporal dependence \cite{lukaszewicz2016navier-stokes}, a property that is inherited by the concentration distribution it produces. {For a schematic example of how turbulence can affect concentration distributions, see Figs.1-5 of Ref.[72] \cite{Richardson}.}  These macroscopic processes are far less well understood than diffusion due to their non-linear nature, but we have shown here that odor tracking strategies on the length scales observed in nature \cite{Lytridis2001} are implausible for a purely diffusive model. 

Although a full description of turbulent behaviour is beyond a purely diffusive model, recurrent attempts have been made to extend these models and approximate the effects of turbulence. These models incorporate time dependent diffusion coefficients \cite{C4CP02019G}, which leads to anomalous diffusion \cite{PhysRevA.35.3081, Cherstvy_2013} and non-homogeneous distributions which may plausibly support odor tracking. {In some cases it is even possible to re-express models using convective terms as a \emph{set} of pure diffusion equations with complex potentials \cite{efficient1,efficient2}}.  Such attempts to incorporate (even approximately) the effects of turbulent air flows are important, for as we have seen (through their absence in a purely diffusive model), this phenomenon is the essential process enabling odors to be tracked.

%\appendix*   % Omit the * if there's more than one appendix.
%
%\section{Uninteresting stuff}
%
%Appendices are for material that is needed for completeness but
%not sufficiently interesting to include in the main body of the paper.  Most
%articles don't need any appendices, but feel free to use them when
%appropriate.  This sample article needs an appendix only to illustrate how 
%to create an appendix.

\begin{acknowledgments}

G.M. would like to thank David R. Griffiths for their helpful comments when reviewing the manuscript as a `professional amateur'. {The authors would also like to thank the anonymous reviewers, whose comments greatly aided the development of this article.} G.M. and D.I.B. are supported by  Army Research Office (ARO) (grant W911NF-19-1-0377; program manager Dr.~James Joseph). AM thanks the MIT Center for Bits and Atoms, the Prostate Cancer Foundation and Standard Banking Group. The views and conclusions contained in this document are those of the authors and should not be interpreted as representing the official policies, either expressed or implied, of  ARO or the U.S. Government. The U.S. Government is authorized to reproduce and distribute reprints for Government purposes notwithstanding any copyright notation herein.

% Anon A would like to thank David R. Griffiths\footnote{not that one} for their helpful comments when reviewing the manuscript as a `professional amateur'.  Anons A and C are supported by  Army Research Office (ARO) (grant XXXXXXXX; program manager Dr.~James Joseph). Anon B thanks the MIT Center for Bits and Atoms, the Prostate Cancer Foundation and Standard Banking Group. The views and conclusions contained in this document are those of the authors and should not be interpreted as representing the official policies, either expressed or implied, of  ARO or the U.S. Government. The U.S. Government is authorized to reproduce and distribute reprints for Government purposes notwithstanding any copyright notation herein.
\end{acknowledgments}

\subsection*{Data Availability}
The data that support the findings of this study are available from the corresponding author
upon reasonable request.
% BEGIN While writing use BibTeX
% \bibliography{refs} % For BibTex
% \nocite{*}
% \bibliography{aipsamp}% Produces the bibliography via BibTeX.

\end{document}